\documentclass[%
  aip,
  jcp,
 amsmath,amssymb,
showpacs,
showkeys,
 reprint,%
]{revtex4-1}
\usepackage[caption=false]{subfig}
\usepackage{graphicx}
\usepackage{epstopdf}
\usepackage{bm}
\usepackage{listings}
\usepackage{amsmath}
\usepackage{ulem}
\usepackage[usenames,dvipsnames,svgnames,table]{xcolor}
\begin{document}


\title[]{Optimization Schemes for Efficient Multiple Exciton Generation and Extraction in Colloidal Quantum Dots}
\author{Fikeraddis A. Damtie}
\email{Fikeraddis.Damtie@teorfys.lu.se}
\affiliation{Mathematical Physics and NanoLund, Lund University, Box 118, 22100 Lund, Sweden.}
\author{Khadga J. Karki}%
\email{Khadga.Karki@chemphys.lu.se}
\affiliation{Chemical Physics and NanoLund, Lund University, Box 124, 22100 Lund, Sweden.}%
\author{T\~{o}nu Pullerits}%
\email{Tonu.Pullerits@chemphys.lu.se}
\affiliation{Chemical Physics and NanoLund, Lund University, Box 124, 22100 Lund, Sweden.}
\author{Andreas Wacker}
\email{Andreas.Wacker@fysik.lu.se}
\affiliation{Mathematical Physics and NanoLund, Lund University, Box 118, 22100 Lund, Sweden.}
\date{25. July 2016: accepted by The Journal of Chemical Physics, 2016}

\begin{abstract}
Multiple exciton generation is a process in which more than one electron hole
pair is generated  per absorbed photon. It allows us to increase  the
efficiency  of solar energy harvesting. Experimental studies
have shown the  multiple exciton generation yield of 1.2 in isolated colloidal
quantum dots. However real photoelectric devices require
  the extraction of electron hole pairs to electric contacts. We provide a
  systematic study of the corresponding quantum coherent processes including
  extraction and injection  and show that a proper design of extraction and
injection rates enhances the yield  significantly up to values around 1.6.
\end{abstract}


\maketitle
\section{\label{sec:Introduction} Introduction}
In recent years colloidal semiconductor quantum dots (QDs) have shown promise
as photovoltaic material.  Quantum confinement in QDs allows  convenient
tuning of the absorption over the whole solar spectrum  leading to the so
called "rainbow" solar cells~\cite{KongkanandJAmChemSoc2008}.  Most promising,
however, is the process known as multiple exciton generation (MEG)
\cite{BeardNL2007,NozikPhysicaE2002,SamburScience2010,SchallerPRL2004,LanNatMater2014,
  BohmNL2015,KamatJPhysChemLett2013,ZhengNanoRes2015,AbdellahJPhysChemC2014}.  MEG is the
generation of more than one electron hole pair with energies close to the
bandgap upon  the absorption of a single high energy photon
~\cite{KarkiNatComm2014,BeardJPhysChemLett2011,
  NozikChemPhysLett2008,ShabaevNL2006,SchallerPRL2004}.  The process is a result of
Coulomb electron-electron interaction (in the form of an inverse Auger
process),  which is more significant in QDs than in bulk structures due to the
forced overlap of electronic wavefunctions~\cite{KlimovARPC2007}.  In
addition, confinement leads to the absence of conservation of momentum,
modified carrier-cooling rates and reduced dielectric screening, all of which
account for enhanced MEG in QDs~\cite{BeardNL2010,KlimovARPC2007}.  Besides
this basic understanding, a detailed microscopic description of MEG in  QDs is
needed for an efficient design and optimization of QD solar cells.  Since the
first demonstration of efficient MEG in PbSe QDs by Schaller and Klimov in
2004~\cite{SchallerPRL2004}, a significant  attention has been paid towards
the study of QD based systems for efficient MEG. Several groups have been
studying MEG efficiency in colloidal semiconductor QDs where they showed a
production  of multiple electron hole pairs upon absorption of a single
photon~\cite{BeardNL2010,BeardJPhysChemLett2011,EllingsonNL2005,BeardAccChemRes2013}.
Lead Chalcogenide QDs (PbS and PbSe), Cadmium Chalcogenide QDs, Indium based
QDs  (InAs and InP) and Silicon QDs are some of the intensively studied QD
systems for exploring the efficiency of MEG~\cite{BeardNL2007,
  GachetNL2010,KarkiSciRep2013,PijpersJPhysChemC2007,SchallerAPL2005, StubbsPRB2010,
  WiseAccChemRes2000}. MEG is commonly measured by ultrafast transient
absorption spectroscopy~\cite{tragerBook2007}, which allows one to capture the
rapid processes of bi-exciton formation and  Auger recombinations on the ps
time scale~\cite{KlimovScience2011}. These processes occur at much faster
timescale as compared to  the lifetimes of single  excitons (ns time
scale)~\cite{CrookerAPL2003}. Recently, we have established a model to study
this time-dependence~\cite{DamtieJPhysConfSer2016}, which describes
bi-exciton formation in good agreement with time-resolved
measurements~\cite{KarkiSciRep2013}.  

In a  real solar cell generating
current, the  extraction and injection of charge carriers is of central
relevance~\cite{ZidekNL2012,ZhengJPhysChemC2014}.  Efficient MEG combined
  with the extraction of electron hole pairs has been demonstrated by several
  groups in the past.  An
  increased peak external quantum efficiency due to MEG was reported for
  PbSe\cite{SemoninScience2011} and PbS \cite{SamburScience2010}  QDs.
  Similarly, increases exceeding $120\%$ have been reported in PbTe QDs
  \cite{BohmNL2015}  and PbSe nanorods \cite{DavisNatComm2015}.
  Quantitatively, the injection and extraction rates of carriers between the
  dot and its environment can depend highly on material parameters such as
the band alignment and the geometry of the physical
realization~\cite{HansenJPhysChemLett2014}.  In this work, we explore the parameter
regimes for efficient extraction of charge carriers produced  with the MEG
scheme. The aim is to guide the design of optimal couplings between the QDs
and the relevant donor and acceptor reservoirs.

For the exciton generation, we consider a short laser pulse. This is guided by
corresponding  optical measurements. The double exciton generation is due to
the Coulomb electron-electron interaction, which we take into account by
diagonalising the QD system with full inter-particle interaction as described
in Ref.~\onlinecite{DamtieJPhysConfSer2016}. Ideally, in the absence of
coupling to the outside environment,  one expects that amplitudes in  the two
states can oscillate indefinitely in a way similar to the Rabi oscillation in
isolated two level system \cite{ShabaevNL2006,EllingsonNL2005}. However, in
reality relaxation and dephasing limit such coherent behavior. This is
described by a Lindblad master equation here. As in
Ref.~\onlinecite{DamtieJPhysConfSer2016}, we take into account pure dephasing,
relaxation within the bands and recombination between the bands, which is
typically the slowest process. In addition, we consider the extraction and
injection of  charge carriers to and from the QD. This allows for a more
realistic description of the yield for the device, which here is defined as
the number of extracted  electrons per absorbed photon. 

  The observable yield depends on two issues: Firstly, the double
  excitons must be produced and secondly their presence needs to be detected. As
  the exciton generation and their subsequent evolution is a coherent
  process, any detection may substantially modify the behavior. Therefore a
  careful definition of the yield based on the experimental setup is important
  \cite{BinksPhysChemChemPhys2011}.
For isolated dots, the yield was measured by the bleach signal of the exciton
  absorption on a timescale of tens of ps after the
  excitation\cite{KarkiSciRep2013}. This average exciton occupation
  probability corresponds to the recombination of excitons, which was used to define the
  yield in our earlier study\cite{DamtieJPhysConfSer2016}. In this work we
  consider a photoelectric device by including the  injection and extraction
  of carriers. This allows for a more practical definition of the yield 
based on the number of the extracted electron hole pairs. 
Compared to our earlier study,
  we find higher yields, as the
extraction reduces the probability  for the double exciton to return back to
the single exciton via Auger recombination.\footnote{The double exciton formation and the Auger recombination are parts of the
same coherent process generated by the Coulomb coupling between these states.
While such terms help in the discussion and qualitative understanding of 
results, one cannot not separate these processes entirely.} This clearly demonstrates the
relevance of properly designed contacts for the injection and the extraction
of the charges and this work contributes to their optimization.

\section{\label{sec:Model and Method} Model}
\begin{figure}
\includegraphics[width=0.95\linewidth]{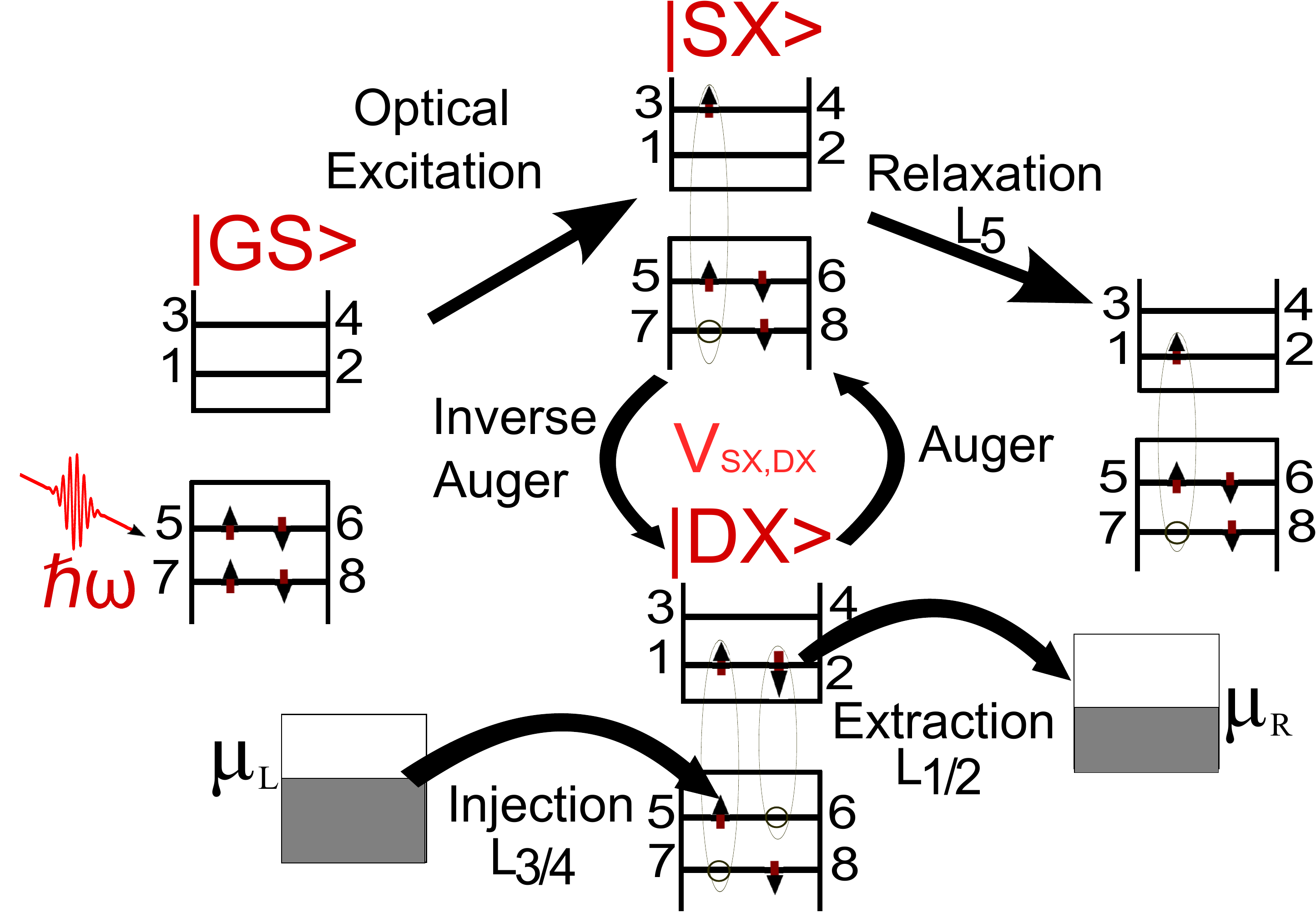}
\caption{The various physical processes considered in the study  namely, optical excitation 
by a resonant pulse, Auger recombination, extraction and injection
to and from a reservoir with chemical potential $\mu_R$ and $\mu_L$ respectively.
In addition we consider relaxation,  which is competing with the inverse Auger process.}
\label{Fig:Process}
\end{figure}
Figure~\ref{Fig:Process} sketches our model for the QD and the
main physical processes considered in the study which are outlined in detail below. 
An optical excitation by a resonant pulse
creates the single exciton state $|SX\rangle$ from the  ground state
$|GS\rangle$. The single exciton state is transferred to the 
double exciton state  $|DX\rangle$ by an inverse Auger process. However,
electron relaxation by other processes, such as phonon scattering, are
 competing process here. 
Once the double exciton is formed,
it is desirable to extract the electrons from the  levels $1$ and $2$
by an efficient mechanism before it returns back to the single exciton
state via an Auger process. In the many-particle description applied here the Auger process 
and its inverse appear as the coherent oscillation between the states $|SX\rangle$ and $|DX\rangle$ 
due to the coupling by the Coulomb matrix element $V_{SX,DX}$, see Ref.~\onlinecite{DamtieJPhysConfSer2016} 
and ~\onlinecite{EllingsonNL2005}
\subsection*{Many particle states}
The total Hamiltonian for the system can be divided into two main parts, the time independent 
Hamiltonian $\hat{H}_0$ of the dot, which fully includes the Coulomb interaction $\hat{H}_{ee}$ 
between the electrons, and the time dependent interaction with the oscillating electric field, 
$\hat{H}_I(t)$. 
\begin{equation}
\mathrm{\hat{H}}_{\mathrm{eff}}(t)=\underbrace{\sum_i E_i \hat{a}_i^\dag\hat{a}_i +\hat{H}_{ee}}_{\hat{H}_0}+\hat{H}_I(t)
\label{Eq:Total Hamiltonian}
\end{equation}
All the operators are expressed in occupation number representation with the creation/annihilation operators 
$\hat{a}_i^\dag/\hat{a}_i $
for electrons in the single particle levels $i$ having energy $E_i$.
The many body states used in our calculations 
are obtained by exact diagonalization of  $\hat{H}_0$. 
We use parameters corresponding to a 4 nm PbS QD following 
Ref.~\onlinecite{DamtieJPhysConfSer2016}, where further details can be found.

\subsection*{Equation of motion for the density matrix}
The QD is excited by an optical pulse, 
which we describe in dipole approximation as
\begin{equation}
\hat{H}_I(t)=\underbrace{eE_0e^{-t^2/\tau^2}\sin(\omega t)}_{E(t)}
\sum_{mn}z_{mn} \hat{a}_m^\dag\hat{a}_n
\end{equation}
Throughout this study we use $\tau=150$~fs, 
$eE_0 z_{37}\approx-0.317$~meV, and $\hbar\omega\approx4.25$~eV, which corresponds to a $0.035\pi$ pulse 
in resonance with the $7\to 3$ (and $8\to 4$) transition taking into account the Coulomb interaction in the ground state. 
(Thus, slightly different values of $\hbar\omega$ will be be used in calculations with a modified  Coulomb strength below.) 
The time evolution of the reduced density operator for the system is evaluated by the Lindblad equation~\cite{LindbladCMP1976}
\begin{multline}\label{Eq:Lindblad}
\hbar\frac{d}{dt}\hat{\rho}_S(t)=i[\hat{\rho}_S(t),\hat{H}_{\mathrm{eff}}(t)]\\
+\sum_{j=1}^{N_{\mathrm{jump}}}\Gamma_j
\bigg[\hat{L}_j\hat{\rho}_S\hat{L}^{\dag}_j-\frac{1}{2}(\hat{L}^{\dag}_j\hat{L}_j\hat{\rho}_S+
\hat{\rho}_S\hat{L}^{\dag}_j\hat{L}_j)   \bigg]\, .
\end{multline}
Here, the jump operators $\hat{L}_j$ describe
different dissipation processes (with rate $\Gamma_j
/\hbar$), which are restoring the ground state for sufficiently long 
times after the excitation.

\subsection*{Dissipative processes considered}
We use the convention that carriers with up spins $\uparrow$ occupy 
odd numbered and those with down spin $\downarrow$ occupy even numbered 
single particle levels in all the definitions below. The following dissipative processes are taken 
into account in the model cf.~Fig.~\ref{Fig:Process}: 
Extraction from the conduction band edge
\begin{equation*}
 \hat{L}_{1}=\hat{a}_{1\uparrow}\textrm{ and }\hat{L}_{2}=\hat{a}_{2\downarrow}
\qquad\mbox{with strength }\Gamma_\mathrm{Ext}
\end{equation*}
Injection into the valence band edge
\begin{equation*}
 \hat{L}_{3}=\hat{a}^{\dag}_{5\uparrow}\textrm{ and }
\hat{L}_{4}=\hat{a}^{\dag}_{6\downarrow}\qquad
\mbox{with strength }\Gamma_\mathrm{Inj}
\end{equation*}
Relaxation in the conduction band 
 \begin{equation*}
 \hat{L}_{5}=\hat{a}^{\dag}_{1\uparrow}\hat{a}_{3\uparrow} 
+ \hat{a}^{\dag}_{2\downarrow}\hat{a}_{4\downarrow} 
 \qquad \mbox{with strength }\Gamma_\mathrm{Rel}
 \end{equation*}
Relaxation in the valence band
 \begin{equation*}
 \hat{L}_{6}=\hat{a}^{\dag}_{7\uparrow}\hat{a}_{5\uparrow} 
+ \hat{a}^{\dag}_{4\downarrow}\hat{a}_{8\downarrow} 
 \qquad\mbox{with strength }\Gamma_\mathrm{Rel}
 \end{equation*}
Recombination across the band gap
 \begin{equation*}
 \hat{L}_{7}=\hat{a}^{\dag}_{5\uparrow}\hat{a}_{1\uparrow} 
+ \hat{a}^{\dag}_{6\downarrow}\hat{a}_{2\downarrow} 
 \qquad \mbox{with strength }\Gamma_\mathrm{Rec}
 \end{equation*}
Dephasing of all states
 \begin{equation*}
\left.
\begin{array}{l}
\hat{L}_{8}=\hat{a}^{\dag}_{1\uparrow}\hat{a}_{1\uparrow} + \hat{a}^{\dag}_{2\downarrow}\hat{a}_{2\downarrow}\\
\hat{L}_{9}=\hat{a}^{\dag}_{3\uparrow}\hat{a}_{3\uparrow} + \hat{a}^{\dag}_{4\downarrow}\hat{a}_{4\downarrow} \\
\hat{L}_{10}=\hat{a}^{\dag}_{5\uparrow}\hat{a}_{5\uparrow} + \hat{a}^{\dag}_{6\downarrow}\hat{a}_{6\downarrow} \\ 
\hat{L}_{11}=\hat{a}^{\dag}_{7\uparrow}\hat{a}_{7\uparrow} + \hat{a}^{\dag}_{8\downarrow}\hat{a}_{8\downarrow} 
\end{array}\right.
 \qquad \mbox{with strength }\Gamma_\mathrm{Deph}\\
\end{equation*}

The jump operators are defined in such a way that they all conserve
the total spin if the particle number is conserved.  The different
decoherence mechanisms, which are phenomenologically described  in
Eq.~(\ref{Eq:Lindblad}), can be associated to all forms of intrinsic
scattering mechanisms other than electron-electron scattering, which
has already been included in the effective Hamiltonian.

In all the simulations, the dephasing rate $\Gamma_{\mathrm{Deph}}=6$~meV which 
corresponds to $\tau_{\mathrm{Deph}}=0.69$~ps 
is applied.
As the recombination is typically the slowest time scale (unless for very weak 
coupling to the reservoir not considered here), 
we neglect this process in our study and set
$\Gamma_\mathrm{Rec}=0$, which corresponds to $\tau_\mathrm{Rec}=\infty$ 
throughout this work. (Test calculations
showed only very small changes of about $5\%$  for
$\Gamma_\mathrm{Rec}=0.1$~meV with the corresponding rate in time units of 
$\tau_\mathrm{Rec}=41.3$~ps, 
which is fairly large compared to the
typical recombination rates in semiconductor dots.)
The extraction and injection processes require certain energy ranges
for  the incoming and outgoing particles. If the jump operator
corresponds to adding a particle (injection),  the initial system with
particle number $N$ has an energy $E(N)$. The incoming particle
should have an energy such that it enters the valence band levels ($5$
or $6$) depending on the spin. After the jump, the system will have
$N+1$ particles with energy $E(N+1)$. The injecting contact for
such a photovoltaic system has its  electrochemical potential a small
margin $\Delta$ above the highest occupied level in the valence band,
as depicted in Fig.~\ref{Fig:Process}. Thus, electrons can only  enter
if $\big(E(N+1)-E(N)\big)\le \big(E_g(4)-E_g(3)\big)+ \Delta$.  Here,
$E_g(N)$ denotes the ground state for $N$ particles as obtained from
the diagonalization of $\hat{H}_0$. Temperature broadening is neglected
for simplicity. Similarly, the jumps associated with the removal
of a particle (extraction), require empty states in  the
corresponding reservoir, which are available  above its electrochemical
potential. This is a small margin $\Delta$ below the lowest level in
the conduction band and  provides the required energy
$\big(E(N)-E(N-1)\big)\ge\big(E_g(5)-E_g(4)\big)-\Delta$ for the
removal of an electron from an $N$-electron state. In all our
calculations, we use $\Delta=0.2$ eV.

\section{\label{sec:Results and Discussion} Results and Discussion}
\begin{figure}[t]
 \subfloat[Extraction rate][Number of extraction  $\Gamma_{\mathrm{Rel}}=3.3$~meV]{
   \includegraphics*[width=0.47\linewidth]{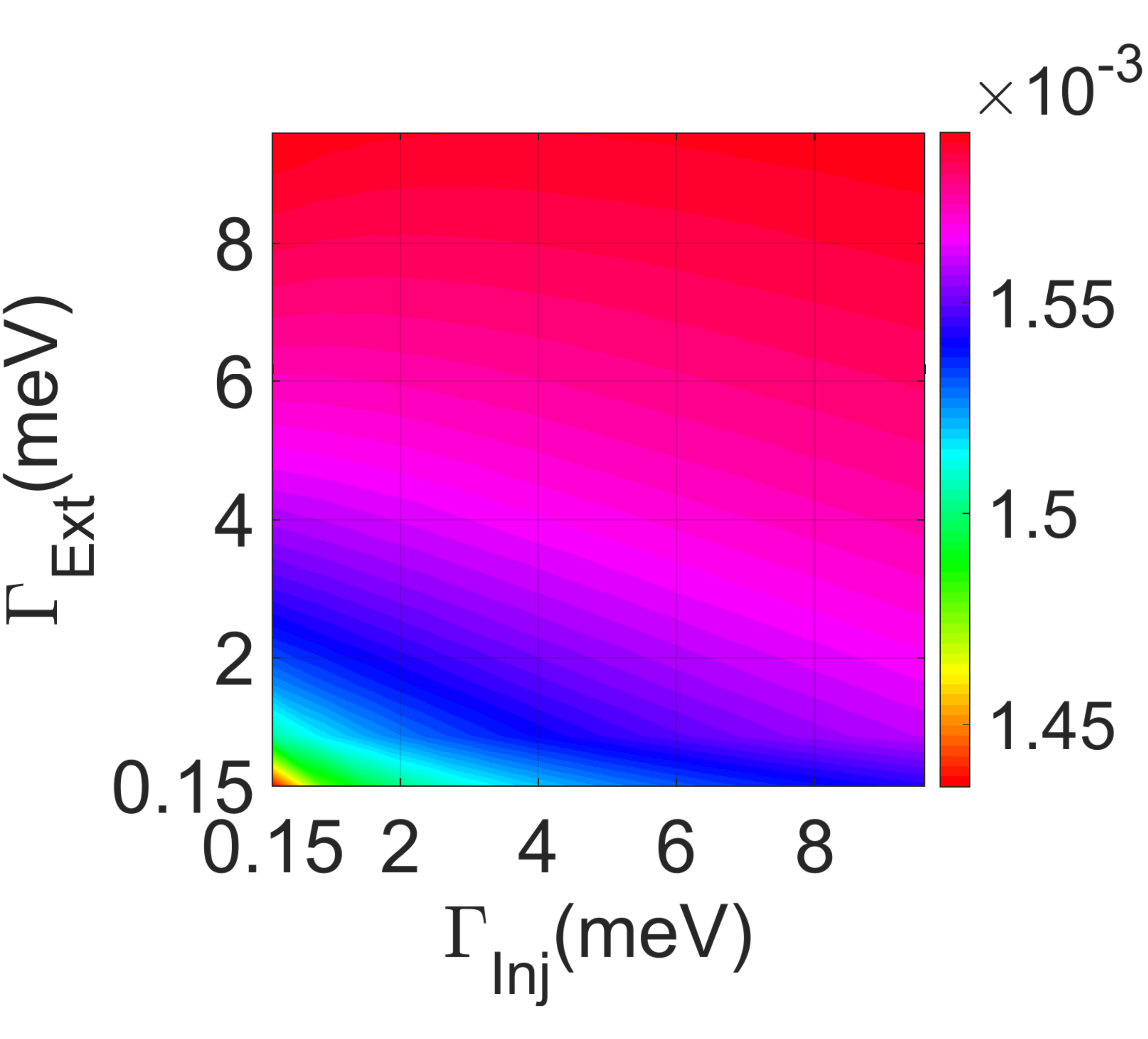}
   \label{SubFig:N_Ext_3_3}
}
 \subfloat[Absorbed energy][Absorbed energy  $\Gamma_{\mathrm{Rel}}=3.3$~meV]{
   \includegraphics*[width=0.47\linewidth]{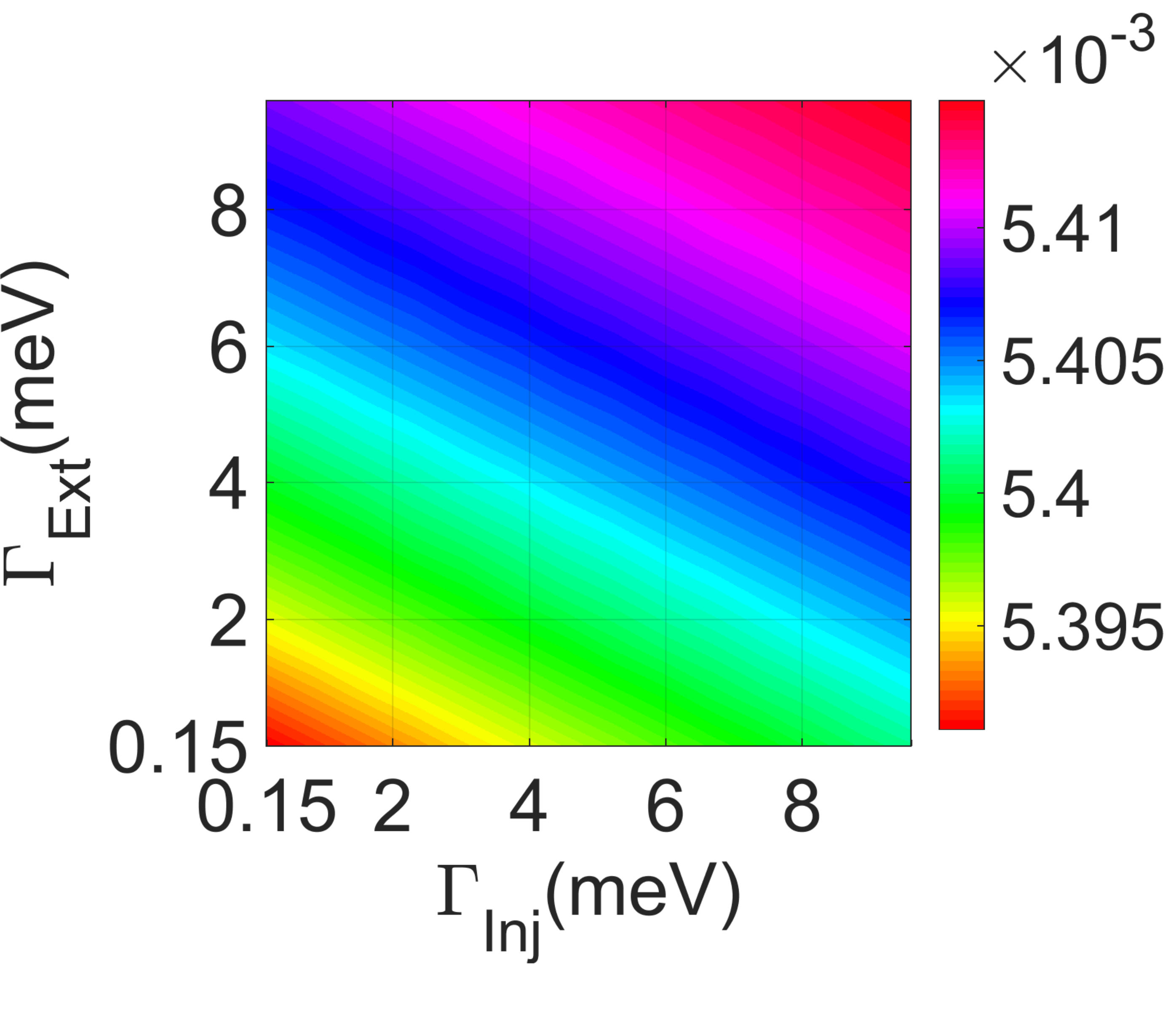}
   \label{SubFig:Abs_Energy_3_3}
 }\\
 \subfloat[Yield][Yield  $\Gamma_{\mathrm{Rel}}=3.3$~meV]{
   \includegraphics*[width=0.47\linewidth]{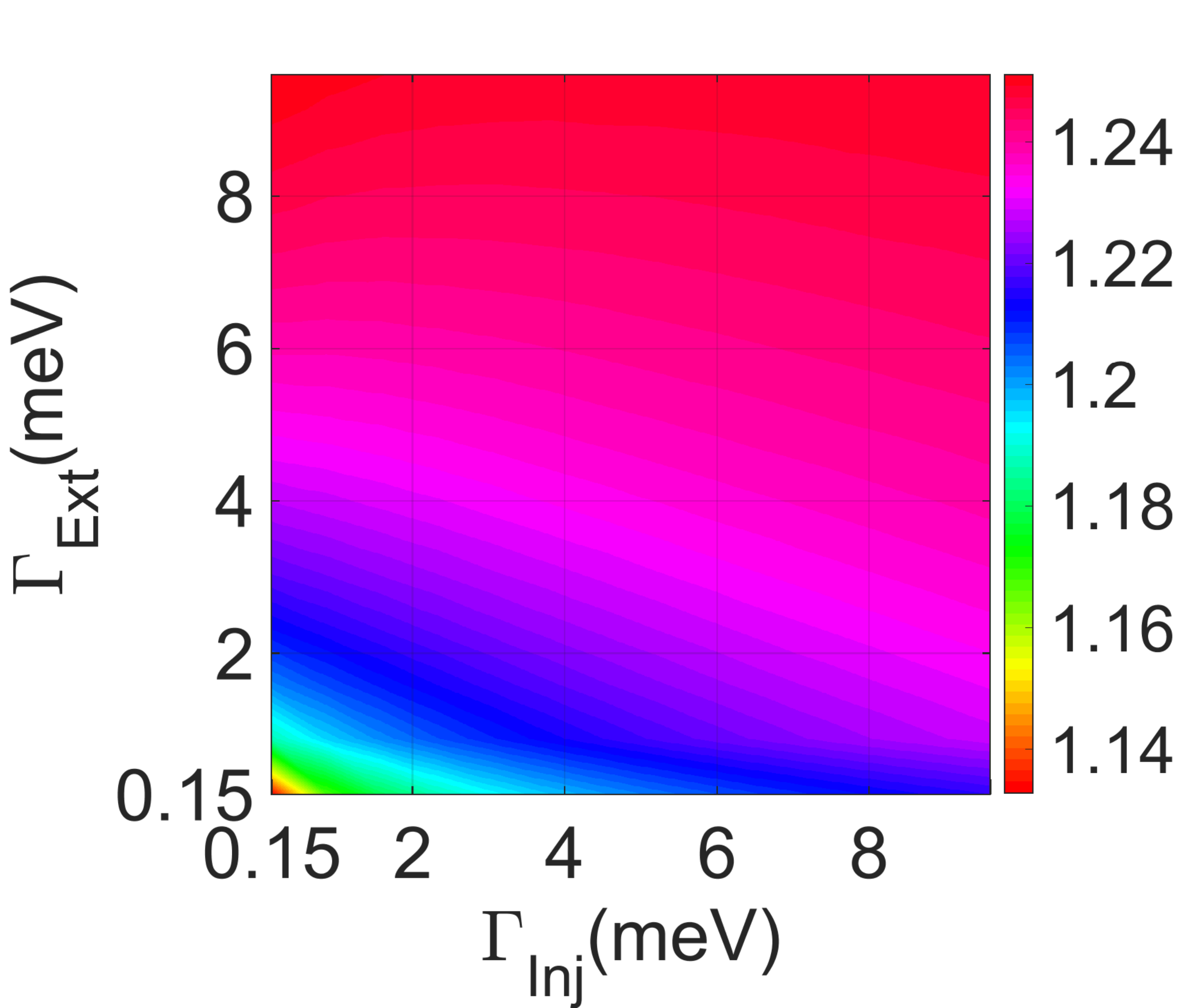}
   \label{SubFig:Yield_Rel_3_3}
 }
 \subfloat[Yield][Yield $\Gamma_{\mathrm{Rel}}=1$~meV]{
   \includegraphics*[width=0.47\linewidth]{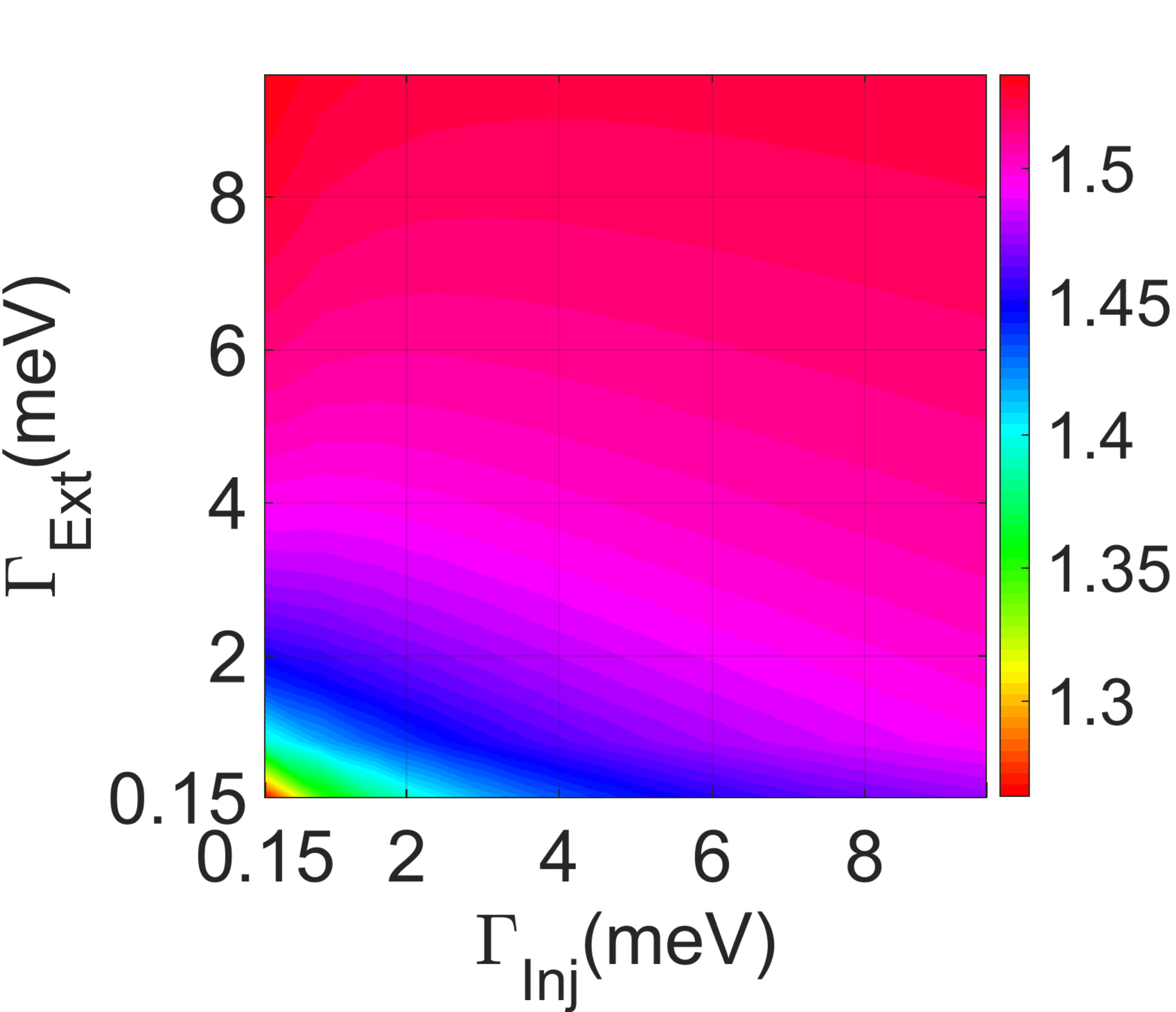}
   \label{SubFig:Yield_Rel_1}
 }
\caption[]{Number of extractions (a) and absorbed energy (b) for different reservoir 
coupling strengths. The ratio between these numbers determines the yield in panel (c). 
Panel (d) shows the yield for a reduced relaxation rate.}
\label{Fig:Fig_3}
\end{figure}

The focus of our work is to determine parameter regimes 
in which the total number of extraction of charged particles 
from the band edges is optimal per single absorbed photon.
Here, we calculate the average number of particles extracted  from the 
conduction band after the pulse excitation. The extraction rate is 
\begin{equation}
\mathrm{Ext(t)}=\sum_{i=1,2}\Gamma_{\mathrm{Ext}}Tr\{\hat{L}_{i}\hat{\rho}\hat{L}_{i}^{\dag}\}
\end{equation}
so that the average number of extracted electron is given by
\begin{equation}
\mbox{Number of extraction}=\int_{-\infty}^{\infty} dt\, \mathrm{Ext}(t).
\end{equation}
This quantity is plotted in 
Figure~\ref{SubFig:N_Ext_3_3} as a function of the extraction rate, $\Gamma_\mathrm{Ext}$, 
for electrons from the conduction band and the injection rate,  
$\Gamma_\mathrm{Inj}$, for electrons into the valance band by appropriately designed contacts. 
As expected, we find that the number of extractions increases with increasing 
reservoir coupling for either contact, as competing relaxation processes become less relevant.
Note that the number of injections, as obtained by summing over jump processes 3 and 4, equals the 
number of extractions, as the system returns into the ground state with 4 electrons occupying the 
levels 5--8 for large times. Thus, the number of extractions and the number of injections constitute 
the same measure for the current flow through the dot.

In order to determine the  yield, we need the number of absorbed photons for comparison. 
Therefore we consider the energy balance for the interaction with the light field \cite{MukamelBook1995}
\begin{equation}
\mathrm{P(t)}=\frac{d }{dt}\langle \hat{H}_\mathrm{eff}(t)\rangle 
=\left\langle {\frac{\partial \hat{H}_{I}(t)}{\partial t}}\right \rangle =e\langle \hat{z}\rangle  \dot{E}(t)
\end{equation}
such that the total energy transferred from the light pulse to the dot is 
\begin{equation}
\mbox{Absorbed Energy}=\int_{-\infty}^t dt\,P(t).
\end{equation}
This quantity is displayed in Figure~\ref{SubFig:Abs_Energy_3_3} and is only changing slightly with the 
contact couplings. Our main point of interest is the yield, i.e., the ratio 
between the number of extraction and the absorbed energy per the incoming photon energy.:
\begin{equation}
\textrm{Yield}=\dfrac{\mbox{Number of extraction}}{\mbox{Absorbed Energy}}\times \mathrm{\hbar\omega_{pulse}}. 
\end{equation}
Figure~\ref{SubFig:Yield_Rel_3_3} shows that
the yield varies between $\approx1.13-1.25$ for different  injection and extraction rates for 
$\Gamma_{\mathrm{Rel}}=3.3$ meV. 
It can be seen that higher rates of extraction and injection in general result in a higher yield. 
This is due to the fact that the double exciton is efficiently extracted before it goes back to the 
single exciton state via the inverse Auger recombination. Comparing the increment in the yield 
as a function of the extraction and injection rate, it can be seen that the yield increases faster 
as a function of extraction and saturates more quickly than upon varying the injection rate. 
The reason for this behavior is that the extraction involves the electrons in the conduction band, which are created 
as a result of the Coulomb electron-electron interaction. On the other hand, the injection rate involves 
the electrons in the valence band, which increases the yield only if the extraction rate is small.
This increase is due to the fact that injection into the level $6$ hinders the Auger process converting the
 $|DX\rangle$ state back to the $|SX\rangle$ state, see Fig.~\ref{Fig:Process}. This provides a relevant 
termination process for the coherent oscillations between the  $|DX\rangle$ and  $|SX\rangle$ state, if the extraction rate is small.
Figure~\ref{SubFig:Yield_Rel_1} shows similar 
results but with a smaller relaxation rate. It can be seen that the overall yield increases
for the ranges shown. Low relaxation rate indeed results in a larger chance for the inverse Auger process to occur
before the single exciton relaxes to some other low energy state. As a result, the creation and extraction of the
double exciton is enhanced for reduced relaxation.    

Since MEG involves a competing process between the Auger kinetics and the relaxation,
it is of interest, to study how these relate to each other quantitatively.
The inverse Auger process is dominated by the Coulomb matrix element $V^{ee}_{1263}=-0.6$ meV.
In order to quantify its role, we now modify this matrix element (as well as the one with spins exchanged and the adjoint ones)
by multiplying them with a factor $F_C$.

\begin{figure}
\includegraphics [scale=.17]{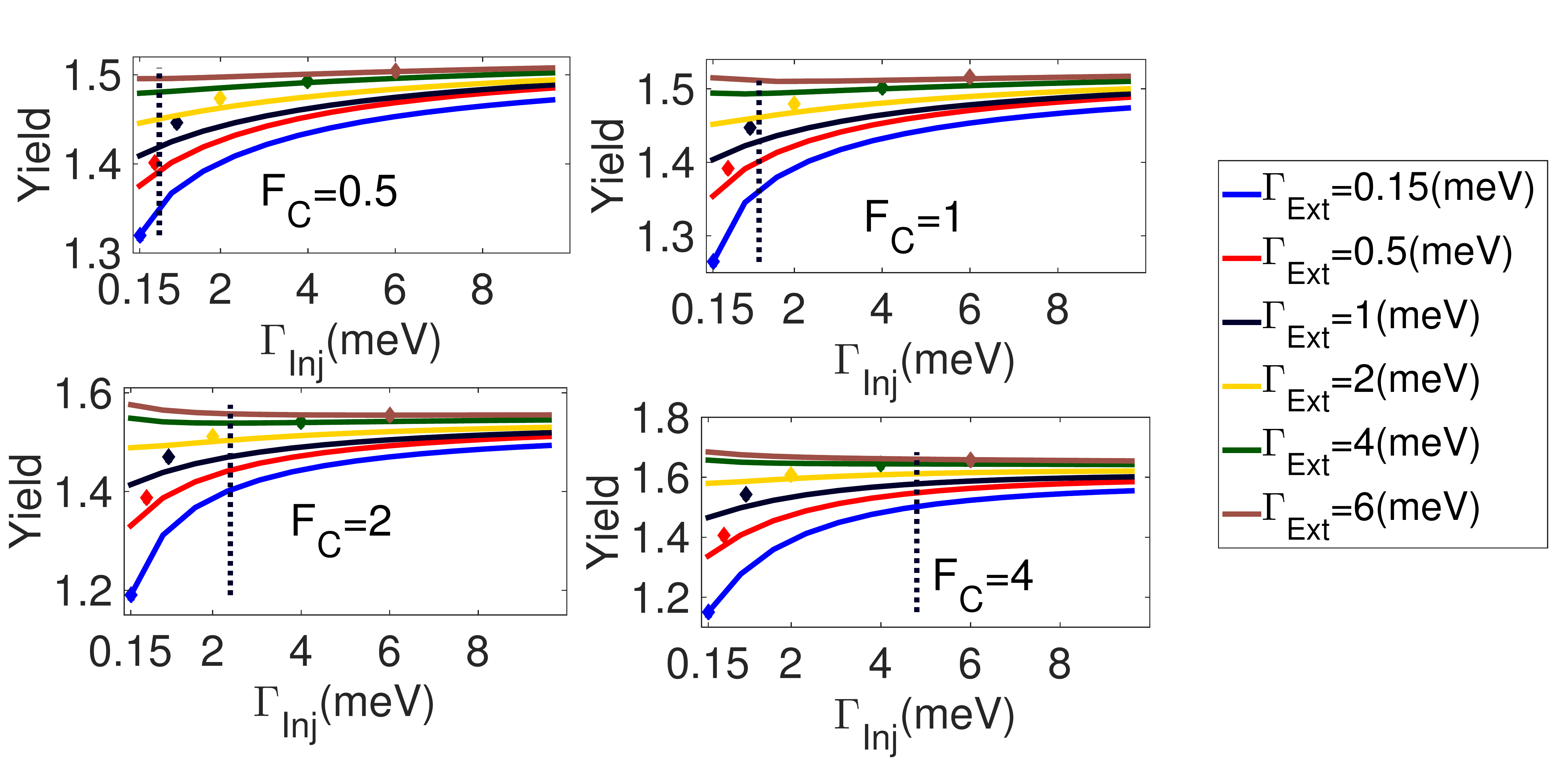}
\caption{Yield as a function of the injection rates for different rates of extraction.
Different strengths for the Auger coupling rate are applied in each panel. 
The vertical line denotes $\Gamma_\mathrm{Inj}=2V^{ee}_{1263}$ and the diamonds refer to the points
$\Gamma_\mathrm{Inj}=\Gamma_\mathrm{Ext}$. The relaxation rate $\Gamma_{\mathrm{Rel}}=1$~meV is used here.
}
\label{Fig:Yield_Coulomb}
\end{figure}

Fig.~\ref{Fig:Yield_Coulomb} shows the yield as a function of injection rate for the different strengths of the 
Coulomb coupling for the Auger terms. 
The energy splitting $|E_{SX}-E_{DX}|\approx 2|V^{ee}_{1263}|$ in each simulation 
is indicated by a dashed vertical line for better comparison with respect to the extraction and 
injection rates. In addition, the extraction rates for each line is indicated by a diamond with the 
corresponding colors for each extraction rate. As expected, the yield increases with increasing either the 
extraction or the Auger coupling for all cases. Here, we find that 
in the case where the Auger Coulomb coupling is dominant over the extraction rate, $2|V^{ee}_{1263}|>\Gamma_{\mathrm{Ext}}$, 
the yield is small as the coherent oscillation between the $|SX\rangle$ and $|DX\rangle$ states is only 
damped by the relaxation processes. 
However, the $|DX\rangle$ state can be conserved by the injection of an electron into the 
valence band and thus we find a significant increase of yield with $\Gamma_\mathrm{Inj}$ under these conditions, 
which levels of for  $\Gamma_\mathrm{Inj}\gg 2|V^{ee}_{1263}|$.  On the other hand,
for $\Gamma_{\mathrm{Ext}}> 2|V^{ee}_{1263}|$ where extraction is dominant, 
the variation of the yield as a function of the injection rate is small. In this case, 
the extraction is sufficient to guarantee high yield.

\section{Conclusion}
In this work, we have examined the conditions for optimal MEG by 
impact ionization upon optical excitation by high energy photons. We have focused on the 
extraction and injection mechanisms 
of charge carriers, which are key ingredients in a realistic device. We have shown 
that an optimal yield can be achieved by an efficient extraction mechanism, which exceeds the 
Coulomb coupling matrix element for the inverse Auger process. 
Furthermore, relaxation should be kept slow. For small extraction rates compared to the 
Coulomb matrix element between the single exciton and double exciton states, 
an increase in the injection rate improves the yield by 
altering the oscillation between $|SX\rangle$ and $|DX\rangle$. 
More importantly, our work shows that the MEG yield in photovoltaic devices can be higher 
than in QDs dispersed in solution that have no contacts for the extraction of the charges. 
\section{Acknowledgments}
We acknowledge the Knut and Alice Wallenberg foundation (KAW), NanoLund as well as 
the Swedish Research Council (VR) for financial support.
\section{References}  

%

\end{document}